# Epitaxial Growth of Ultraflat Bismuthene with Large Topological Band Inversion Enabled by Substrate-Orbital-Filtering Effect

Shuo Sun,[†, ‡, □] Jing-Yang You,[‡, □] Sisheng Duan,[‡, □] Jian Gou,[‡] Yong Zheng Luo,[∥] Weinan Lin,[††] Xu Lian,[§] Tengyu Jin,[‡, ‡‡] Jiawei Liu,[‡] Yuli Huang,[‡, ‡‡] Yihe Wang,[§, ‡‡] Andrew T. S. Wee,[‡] Yuan Ping Feng,[‡] Lei Shen,*[, ∥] Jia Lin Zhang,*[, #] Jingsheng Chen,*[, †] and Wei Chen*[, ‡, §, ‡‡, §§]

[†]Department of Materials Science and Engineering, [‡]Department of Physics, [§]Department of Chemistry and [∥]Department of Mechanical Engineering, National University of Singapore, 119077, Singapore

[#]School of Physics, Southeast University, Nanjing 211189, China

[††]Department of physics, Xiamen University, Xiamen 361005, China

[‡‡]Joint School of National University of Singapore and Tianjin University, International Campus of Tianjin University, Binhai New City, Fuzhou 350207, China

[§§]National University of Singapore (Suzhou) Research Institute, Suzhou, 215123, China

[□]These authors contributed equally: Shuo Sun, Jing-Yang You and Sisheng Duan

***Corresponding Authors**: L. S., (shenlei@nus.edu.sg) J. L. Z., (phyzjl@seu.edu.cn) J. C. (msecj@nus.edu.sg) and W. C. (phycw@nus.edu.sg).

ABSTRACT: Quantum spin Hall (QSH) systems hold promises of low-power-consuming spintronic devices, yet their practical applications are extremely impeded by the small energy gaps. Fabricating QSH materials with large gaps, especially under the guidance of design principles, is essential for both scientific research and practical applications. Here, we demonstrate that large on-site atomic spin-orbit coupling can be directly exploited *via* the intriguing substrate-orbital-filtering effect to generate large-gap QSH systems and experimentally realized on the epitaxially synthesized ultraflat bismuthene on Ag(111). Theoretical calculations reveal that the underlying substrate selectively filters Bi $p_z$ orbitals away from the Fermi level, leading $p_{xy}$ orbitals with nonzero magnetic quantum numbers, resulting in large topological gap of ~1 eV at the K point. The corresponding topological edge states are identified through scanning tunneling spectroscopy combined with density functional theory calculations. Our findings provide general strategies to design large-gap QSH systems and further explore their topology-related physics.

KEYWORDS: *quantum spin Hall effect, substrate-orbital-filtering effect, ultraflat bismuthene monolayer, STM, spin-orbit coupling*

**INTRODUCTION**

Two-dimensional (2D) topological materials (such as the quantum spin Hall (QSH) insulators), which exhibit one-dimensional topological edge states within the bulk band gap, are potential candidates to realize dissipationless transport. With graphene as the initially predicted 2D topological insulator to possess the QSH effect,[1,2] many other 2D topological materials have also been discovered.[3–13] However, most of them feature very small energy gap, *e.g.*, graphene (0.8 × $10^{-3}$ meV),[1,14] silicene (2 meV),[8] germanene (24 meV)[8] and stanene (100 meV),[10] and hence extremely low temperatures[6,7] are usually required to suppress the thermal fluctuation, which restricts their practical applications. Therefore, fabricating QSH systems with large gaps is of great value for both scientific research and practical applications.

Guided by the Kane-Mele model,[1] one general approach for achieving large nontrivial gap is to introduce heavy elements into the 2D material to considerably enhance the spin-orbit coupling (SOC). It is worth noting that the experimental synthesis of 2D topological materials requires the supporting substrates. The accompanying interactions between the 2D material and its substrate can considerably affect the topological properties of the 2D material. Nevertheless, the interfacial interactions could also be employed as a promising pathway to tailor the topological properties of 2D materials, *e.g.*, by strains[15–17] or by moiré superstructures.[18,19] Through theoretical calculations, Zhou and Liu *et al.* proposed an intriguing substrate-orbital-filtering effect,[20] where the interaction with the substrate could select the orbital composition around the Fermi level, converting the system from a topologically trivial phase into a nontrivial phase. It is worth noting that the substrate-orbital-filtering effect has been also applied to generate larger QSH gap (~0.5–0.6 eV) in the very recent theoretically proposed system of plumbene on the BaTe(111) surface.[21] Therefore, it is expected that the combination of these two strategies, *i.e.*, bringing in the heavy elements and substrate-orbital-filtering effect, can produce a large-gap 2D topological material.

Bismuth is a heavy element (Z = 83) with strong SOC. Its 2D form, known as bismuthene with buckled honeycomb lattice, is considered as a representative 2D topological insulator possessing extraordinarily large nontrivial gap,[20,22–28] and has stimulated significant experimental investigations,[29–32] wherein the most significant advancement is the report of bismuthene on a SiC(0001) substrate hosting the largest topological gap (0.8 eV) revealed by scanning tunneling spectroscopy (STS)[33] combined with detailed theoretical paradigm.[34] Although bismuthene has been successfully synthesized on different substrates, in many cases, the unavoidable strong interfacial interactions from the underlying substrates and the complex energy bands near the Fermi level impede the observation of the topological states in the experiment. Utilizing the substrate-orbital-filtering effect to modulate the topological states and achieve large nontrivial energy gap has been rarely explored experimentally.

Here, we report the realization of bismuthene monolayer with ultraflat honeycomb lattice on Ag(111) by molecular beam epitaxy (MBE). The atomic structure and topological edge states of the synthesized ultraflat bismuthene are clearly identified at the atomic scale through low temperature scanning tunneling microscopy/spectroscopy (LT-STM/STS) combined with density functional theory (DFT) calculations. Detailed analysis suggests that the underlying substrate not only plays a critical role in stabilizing the ultraflat structure, but also provides substrate-orbital-filtering effect in this ultraflat bismuthene, which transforms the otherwise topologically trivial freestanding ultraflat bismuthene into a nontrivial phase with large topological gap of ~1.0 eV at the K point, which is six orders of magnitude higher than that of freestanding graphene. Our findings of ultraflat bismuthene enrich the rarely explored field of planar graphene-like materials and provide a general mechanism to design large-gap QSH systems for room-temperature applications.

**RESULTS AND DISCUSSION**

**Epitaxial Growth of Ultraflat Bismuthene Monolayer on Ag(111)**

Previous study has demonstrated that the surface structure of bismuth deposited on room-temperature Ag(111) varies as a function of the Bi coverage. At low coverage, ordered ($\sqrt{3} \times \sqrt{3}$) R30° BiAg$_2$ alloy with enhanced Rashba spin-orbit splitting is formed.[35] Further increasing the coverage above 0.55 monolayer, the alloying-to-dealloying transition occurs and an ordered Bi (*p*

× √3) phase is observed.[36] Beyond 1 monolayer, Bi(110) bilayer ribbons with black phosphorus-like puckered structures develop on the surface.[36] Intriguingly, we found that the growth behavior of Bi on low temperature Ag(111) is different. Figure 1a is the schematic diagram of the epitaxial process of Bi on Ag(111). The Bi atoms were evaporated from a crucible heated at 480 °C and deposited onto Ag(111) held at 200 K for 3 min in ultrahigh vacuum (UHV) condition, leading to the formation of uniform and continuous 2D Bi films (Figure 1b). The height profile measured along the white line in Figure 1b reveals the formation of Bi monolayer with a height of 1.6 Å, which is smaller than the reported value for buckled honeycomb bismuthene (4.5 Å).[18] Figure 1c shows the atomic resolution STM image of the Bi films, where the honeycomb arrangement of the Bi atoms is clearly resolved. Intriguingly, our synthesized honeycomb structure is completely flat with zero buckling height, namely the ultraflat bismuthene, which is different from the reported buckled honeycomb structures with Bi $sp^3$ hybridization.[18] The lattice constant of the ultraflat bismuthene measured is around 5.7 Å (Figure 1d), perfectly matched with a 2 × 2 supercell of the underlying Ag(111) surface (2.88 Å) with only 1% lattice mismatch. Figure 1e shows the structure model of the ultraflat bismuthene on Ag(111), with every Bi atoms located at the hollow sites of Ag(111). The simulated STM image based on this model perfectly reproduces the experimental observation (Figure 1f), confirming the formation of a fully planar honeycomb bismuthene monolayer on Ag(111).

Figure 1g shows the calculated total energy per Bi atom as a function of the buckling degree for both freestanding and Ag(111)-supported bismuthene. In both cases, zero buckling corresponds to the most stable configuration, while the inclusion of the Ag(111) surface can dramatically decrease the energy by 0.76 eV/atom. The decrease of the energy is attributed to the interaction between the Bi atoms and the underlying Ag(111) surface. By increasing the buckling height, some of the Bi atoms decouple from the substrate, leading to a gradual increase of the total energy and finally to an unstable geometry. It is noted that the freestanding bismuthene is unstable in both zero and higher buckling height, indicating a critical role played by the substrate in stabilizing the monolayer bismuthene (see detailed discussion in Supporting Information, Figure S1). The stabilization arises from the formation of strong bond between the Bi atoms and the Ag atoms. Figure 1h demonstrates a calculated cross-sectional electron localization function (ELF) pattern of the ultraflat bismuthene on Ag(111), which is along the black arrow in Figure 1e. The electrons highly localized at the Bi−Bi pair, and the ELF value between the Bi atoms and the nearest underlying Ag atom (marked as red) is relatively high, indicating a strong bonding between them. Further calculations reveal that the strong coupling between the Ag $s$ orbitals and the Bi $p_z$ orbitals induces such electron clouds uneven distribution (depicted with the black dashed line in Figure 1h), which not only stabilizes the structure of the ultraflat bismuthene but also provides the substrate-orbital-filtering effect to generate the large SOC-induced gap, which will be discussed in detail in the following DFT calculations.

**Modulating Topological Electronic States by the Substrate-Orbital-Filtering Effect**

The experimentally realized ultraflat bismuthene monolayer on Ag(111) provides a platform to investigate the topological properties of bismuthene under the SOC and substrate effect. First, we calculate the band structures of freestanding bismuthene, bismuthene on one layer Ag(111) {bismuthene/1L-Ag(111)} and bismuthene on three layer Ag(111) {bismuthene/3L-Ag(111)}, respectively (Figure S2). Notice that in order to reveal the substrate-orbital-filtering effect, the freestanding bismuthene is also calculated (Figure 2a-c) for comparison. Without SOC, four Dirac cones are observed in the freestanding bismuthene (black dashed circles in Figure 2a). Two of them are located at the K point, with the upper and lower Dirac cones consisting of $p_{xy}$ orbitals

(red dotted lines) and $p_z$ orbitals (green dotted lines), respectively. Another two Dirac cones are formed through the crossing of $p_{xy}$ orbitals and $p_z$ orbitals, with the Dirac point located in between the path K-M and K-Γ, respectively. When SOC is presented, a SOC-induced gap appears near the Fermi level, as shown in Figure 2b. However, the topological invariant $Z_2$ of the freestanding bismuthene is zero (see the first-principles calculations in Methods), indicating that the freestanding bismuthene is topologically trivial. Accordingly, we calculate the edge states of semi-infinite freestanding bismuthene with a zigzag-type termination. As shown in Figure 2c, there are two Dirac cones within the band gap, further indicating the topologically trivial for the freestanding bismuthene. It should be noted that the edge states here are unstable with the adiabatic evolutions taken into consideration, which makes the freestanding bismuthene topological trivial. The degenerate points in Figure 2c are occasional and not protected by symmetry (Figure S3).

To investigate the substrate effect on the topological properties of bismuthene, we take one layer Ag(111) into account. Clearly, the strong coupling with the Ag $s$ orbitals pushes the Bi $p_z$ orbitals away from the Fermi level to the conducting bands and shifts the Dirac cone formed by $p_{xy}$ orbitals to the Fermi level (Figure 2d). A large topological gap of ~1 eV at the K point is opened with SOC in Figure 2e, which is about six orders of magnitude larger than that produced in graphene. This is even larger than the originally predicted value (0.8 eV) on Si or SiC substrate,[20,33] because of the larger SOC strength of Ag substrate in our system.[23] It is noted that noticeable Rashba splitting can also be clearly observed due to the breaking of the inversion symmetry.[34,37] Intriguingly, the filtering of $p_z$ orbitals away from the $p_{xy}$ orbitals at the K point through the interaction with Ag(111) surface induces a topologically nontrivial phase, as proved by the calculated topological invariant $Z_2 = 1$ (Figure 2f). The nontrivial $Z_2$ system can exhibit robust topological edge states, wherein more electronic states can be detected in the edge comparing with the bulk area. STS measurements can probe the local electronic density of states (DOS) and enable us the experimental investigations of the topological edge states at the atomic scale, which will be demonstrated below.

To understand the topological nontrivial phase induced by the substrate-orbital-filtering effect, we need to revisit the origin of the 2D topological insulator. Initially, the concept of 2D topological insulator was proposed by the Kane-Mele model,[1,2] in which the SOC opens a small energy gap at the Dirac point, converting the system to a topological phase with a gapped interior and gapless edge states. Kane and Mele proposed that the Dirac band of the honeycomb lattice (graphene) is derived from the $p_z$ orbitals. The SOC in this system is between the next-nearest neighbors hopping with high-order perturbation theory with relatively small strength, thus producing a tiny gap. In our experiment, the Bi $p_z$ orbitals are pushed out of the Fermi level, leaving one Dirac cone formed by $p_{xy}$ orbitals exactly located at the Fermi level. In this case, the low energy physics is attributed to $p_{xy}$ orbitals in the honeycomb lattice.[12,38] The on-site atomic SOC of $p_{xy}$ orbitals is about two orders of magnitude larger than the next-nearest neighbors SOC of $p_z$ orbitals, therefore giving rise to a considerably enhanced topological gap at the K point compared with graphene. As for the freestanding bismuthene, $p_{xy}$ and $p_z$ orbitals are not separated, which could generate two Dirac cones within the gap, leading to a trivial band topology ($Z_2 = 0$).[20]

To establish the electronic structures of the ultraflat bismuthene more accurately, further DFT calculations based on bismuthene/3L-Ag(111) structure were performed. When more Ag(111) layers are taken into account, the $p_z$ orbitals are pushed further away from the Fermi level, while the Dirac band consisting of $p_{xy}$ orbitals is still located at the Fermi level (Figure 2g). An even larger SOC-induced gap of ~1.4 eV at the K point (indicated by the blue arrow) is opened as a result of more efficient substrate-orbital-filtering modulation. The system is still topologically nontrivial with a topological invariant $Z_2 = 1$ (Figure 2i) although the interaction with the Ag(111) substrate introduces undesired electronic states into the gap as shown in Figure 2h.

**Experimental Investigation of the Topological Edge States in Bismuthene on Ag(111)**

In our experiment, both the zigzag edges and armchair edges can be clearly observed (Figure S4), enabling the investigation of their edge states at the same time. Figure 3a shows the *dI/dV* spectra measured at the zigzag edge, armchair edge and the central part of the ultraflat bismuthene island, respectively. Three pronounced peaks located at around -0.42 eV, -0.27 eV and 0.34 eV are observed at the zigzag edge and armchair edge. However, the peak intensity dramatically enhanced at the edges compared with that measured in the interior island (black line in Figure 3a). To assign the origin of each peak, we compared the calculated bulk DOS with the experimental bulk STS (Figure S5). It is found that the position of the calculated DOS is slightly higher (~0.2 eV) than that of the experimental STS. Nevertheless, their peaks and overall shapes are in good agreement. This quantitative discrepancy can be ascribed to the generalized gradient approximation (GGA) in DFT, which has been reported in other systems, *e.g.*, ultraflat stanene on Cu(111).[39] Thus, the pronounced peak around -0.42 eV is derived from the topological edge states residing at near -0.2 eV in Figure 2i, while the peak at -0.27 eV is attributed to the trivial edge states located around the Fermi level in Figure 2i. The peak centered at 0.34 eV above the Fermi level is contributed by the underlying Ag surface (Figure S6). Figure 3c, d is the *dI/dV* line mappings measured across the zigzag edge and armchair edge of the ultraflat bismuthene respectively. The trajectory is indicated by the white arrows pointing from the bare Ag(111) surface (0 Å) to the center (50 Å) of the bismuthene island (Figure 3b). The states at around -0.42 eV indeed show an enhanced intensity at the edges and decay rapidly toward the bulk with a depth of about 11 Å.

We have also compared the topography STM image (Figure 4a) with the corresponding *dI/dV* mapping image at a selected bias of -0.4 V (Figure 4b). Clearly, the edges present an enhanced intensity and decay towards the bulk with a width of about 11 Å, in accordance with *dI/dV* line mapping in Figure 3c, d. The electronic properties of the bismuthene nanoribbons are also investigated theoretically. To simplify the calculations, a bismuthene/1L-Ag(111) structure is adopted. Here, the width of the nanoribbons is set to be ten unit cells (55.8 Å of zigzag and 56.8 Å of armchair), ensuring no interaction between the two side edges. As shown in Figure 4c, d, the edge states (blue clouds) with a width of two unit cells (9.85 Å of zigzag; 11.37 Å of armchair) can be clearly resolved, which is well matched with the experimental measurements (Figure 3c, d, and Figure 4b). Moreover, helical edge states that connecting the valence and conduction band are clearly observed, which is a distinguishing feature of the QSH systems (Figure 4e, f). The spin-splitting of the helical edge states can be observed as well due to the strong Rashba effect in the system. Therefore, the combination of DFT calculations with STS measurements confirms the existence of nontrivial topological edge states in the bismuthene/Ag(111) system, which is derived from the $p_{xy}$ orbitals with on-site atomic SOC enabled by the substrate-orbital-filtering effect.

Fabricating 2D topological insulators with large gaps is vital for the spintronic device applications, in which enhancing SOC effect is one of the most efficient approaches to generate large gaps. Previous studies have demonstrated that using heavier elements than C is a common way to enhance the SOC effect.[8,10] In this circumstance, the SOC-induced gaps are still relatively small because of the minute SOC derived from $p_z$ orbital with the next-nearest neighboring hopping with high-order perturbation theory. On the other hand, these SOC-induced gaps are not guaranteed to be topologically nontrivial (as discussed for the freestanding ultraflat bismuthene in Figure 2a-c). The orbital selection plays a pivotal role, *i.e.*, only separated $p_z$ or $p_{xy}$ orbitals in the honeycomb lattice can give rise to a nontrivial phase. The $p_{xy}$ orbitals with the on-site atomic SOC can dramatically enhance the QSH gap compared with the $p_z$ orbitals[12,38,40] of weak next-nearest neighbors SOC. Therefore, utilizing on-site atomic SOC with heavy elements in honeycomb lattice is bound to generate a large nontrivial gap.

In our system, exploiting the on-site atomic SOC in the ultraflat bismuthene/Ag(111) can generate a large SOC-induced topological gap at the K point, as achieved by the substrate-orbital-filtering effect, where the *s* orbitals of Ag atoms selectively lift the $p_z$ orbitals of Bi atoms away from the Fermi level. This is different from the previous studies on the substrate-supported 2D materials, where the strong interfacial interactions can break the nontrivial topological invariants due to the orbital hybridization with substrate. Here, *via* low temperature MBE method, the structure of ultraflat honeycomb bismuthene on Ag(111) was constructed, where the substrate-orbital-filtering effect comes into play. Instead of breaking the topology, the substrate-orbital-filtering effect leads to a topological phase transition from the topologically trivial in freestanding ultraflat bismuthene into nontrivial. Although the unwanted metallic states from the substrate hinder the observation of the insulating gap in experiment, the good agreement between experimental results and theoretical calculations strongly suggests that the topological edge states still exist and the underlying physics remains. Our findings of large topological band inversion enabled by substrate-orbital-filtering effect can inspire devices designing for the direct transport measurement of the topological edge states,[41] *e.g.*, the universal quantized conductance, in future.

**CONCLUSIONS**

In summary, a facile method for the fabrication of high quality 2D ultraflat bismuthene on Ag(111) *via* low temperature MBE has been demonstrated. The combination of experimental results and theoretical calculations suggests that the on-site atomic SOC can be directly exploited to realize large-gap topological 2D materials, enabled by the decisive role of the substrate in selectively filtering out the zero-magnetic-quantum-number orbitals, such as $p_z$ and $d_z^2$. Our findings can provide a general design principle to fabricate large-gap 2D topological insulators, which could further extend the research of low-dimensional topological physics and devices for room-temperature applications.

**METHODS**

**Sample Preparation and Characterization.** Experiments were performed in an Omicron LT-STM system with a base pressure of $1 \times 10^{-10}$ mbar. Clean Ag(111) substrate was prepared by several cycles of Ar$^+$ sputtering (1.5 kV, $5 \times 10^{-5}$ mbar) for 30 min and subsequent annealing at 400 °C for 30 min. A bismuth source (99.999%) was used as the precursor, which was heated at 480 °C. The ultraflat bismuthene monolayer was prepared by depositing Bi atoms onto Ag(111) held at 200 K for 3 min. Here, the Ag substrate was intentionally kept at a low temperature of 200 K. Otherwise, the higher substrate temperatures will lead to the formation of bismuth-silver alloy and bismuth overlayer with rectangular symmetry (Figure S7). All the STM/STS measurements were carried out at 77 K using a tungsten tip, with the bias voltage applied to the sample. STS was measured by a lock-in technique with a modulated voltage of 10-mV (root mean square) at the frequency of 963-Hz. All the STM images have been processed using WSxM software.[42]

**First-Principles Calculations.** Our first-principles calculations were based on DFT as implemented in the Vienna *ab Initio* Simulation Package (VASP),[43] using the projector augmented-wave method.[44] The GGA with the Perdew-Burke-Ernzerhof[45] realization was adopted for the exchange-correlation functional. The several atomic layers were placed under a vacuum layer of 15 Å. The plane-wave cutoff energy was set to 500 eV. A Monkhorst-Pack k-point mesh[46] with a size of 19 × 19 × 1 was used for the Brillouin zone sampling, except for the nanoribbon calculation with a sampling of 9 × 1 × 1. The crystal structure was optimized until the forces on the ions were less than 0.001 eV/Å, and the total energy was converged to $10^{-8}$ eV with the Gaussian smearing method. Except for the freestanding bismuthene, the interlayer van der Waals interaction was taken into consideration during the calculation of bismuthene on the Ag substrate. The surface spectrum was calculated using the Wannier functions and the iterative Green's function method.[47–50]

## ASSOCIATED CONTENT

## Author Contributions

S.S., J.C. and W.C. conceived the experiments and provided financial support. S.S., J.L.Z., S.D. and J.G. performed the MBE growth and STM/STS measurements. J.Y., Y.Z.L., and L.S. performed theoretical calculations. S.S., J.Y., J.C. and W.C. analyzed the data. S.S., J.L.Z. and W.C. wrote the paper. All authors discussed the results and helped in writing the paper. □These authors contributed equally: Shuo Sun, Jing-Yang You and Sisheng Duan

## Notes

The authors declare no competing financial interest.

## ACKNOWLEDGMENTS

This work was financially supported by Natural Science Foundation of China (U2032147, 21872100), Natural Science Foundation of Jiangsu Province under BK20210199, Singapore MOE Academic Research Fund Tier 2 Grant No. MOE-2019-T2-1-002 and Singapore NRF Grant No.

R-144-000-405-281. L.S acknowledges Singapore MOE Academic Research Fund Tier 1 Grants No. R-265-000-651-114 and No. R265-000-691-114.

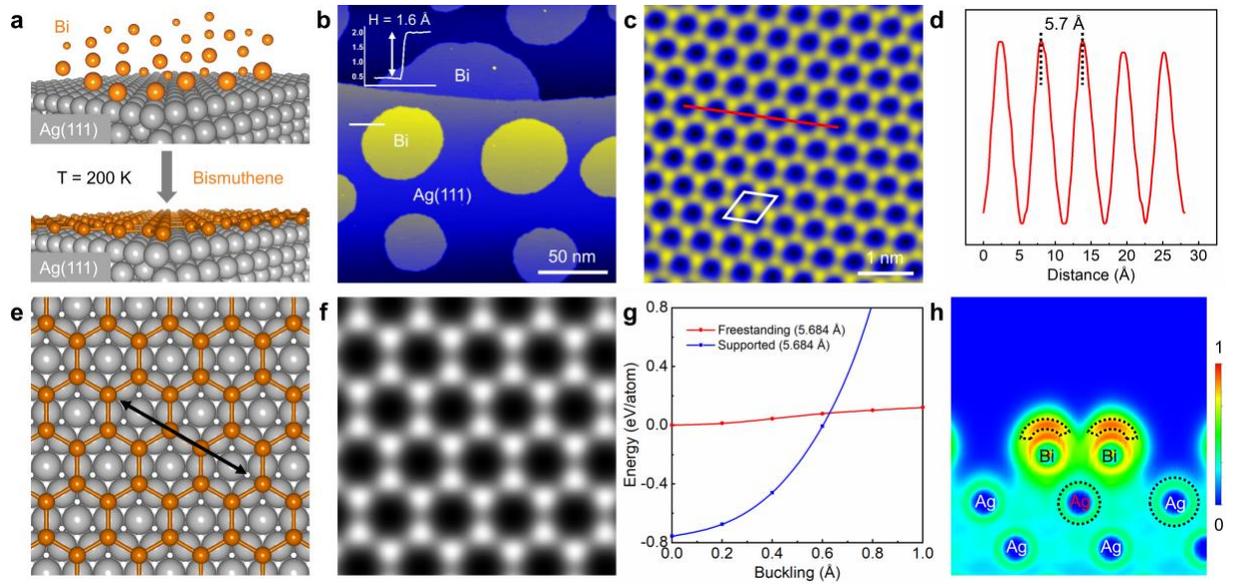

**Figure 1.** Epitaxial growth of ultraflat bismuthene on Ag(111). (a) Schematic diagram of the fabrication process. (b) Large scale STM image of ultraflat bismuthene on Ag(111) ($V_s = 0.5$ V, $200 \times 200$ nm$^2$). Inset, the height profile along the white line in panel (b). (c) Atomic resolution STM image of the bismuthene, where the unit cell is marked by the white rhombus ($V_s = 0.1$ V, $5 \times 5$ nm$^2$). (d) Line profile along the red line in panel (c). (e) Schematic atomic model of the ultraflat honeycomb bismuthene. (f) Simulated STM image based on the model in panel (e). (g) Total energy per Bi atom as a function of the buckling for the freestanding and Ag(111)-supported bismuthene. (h) Cross-sectional ELF along the black arrow in panel (e).

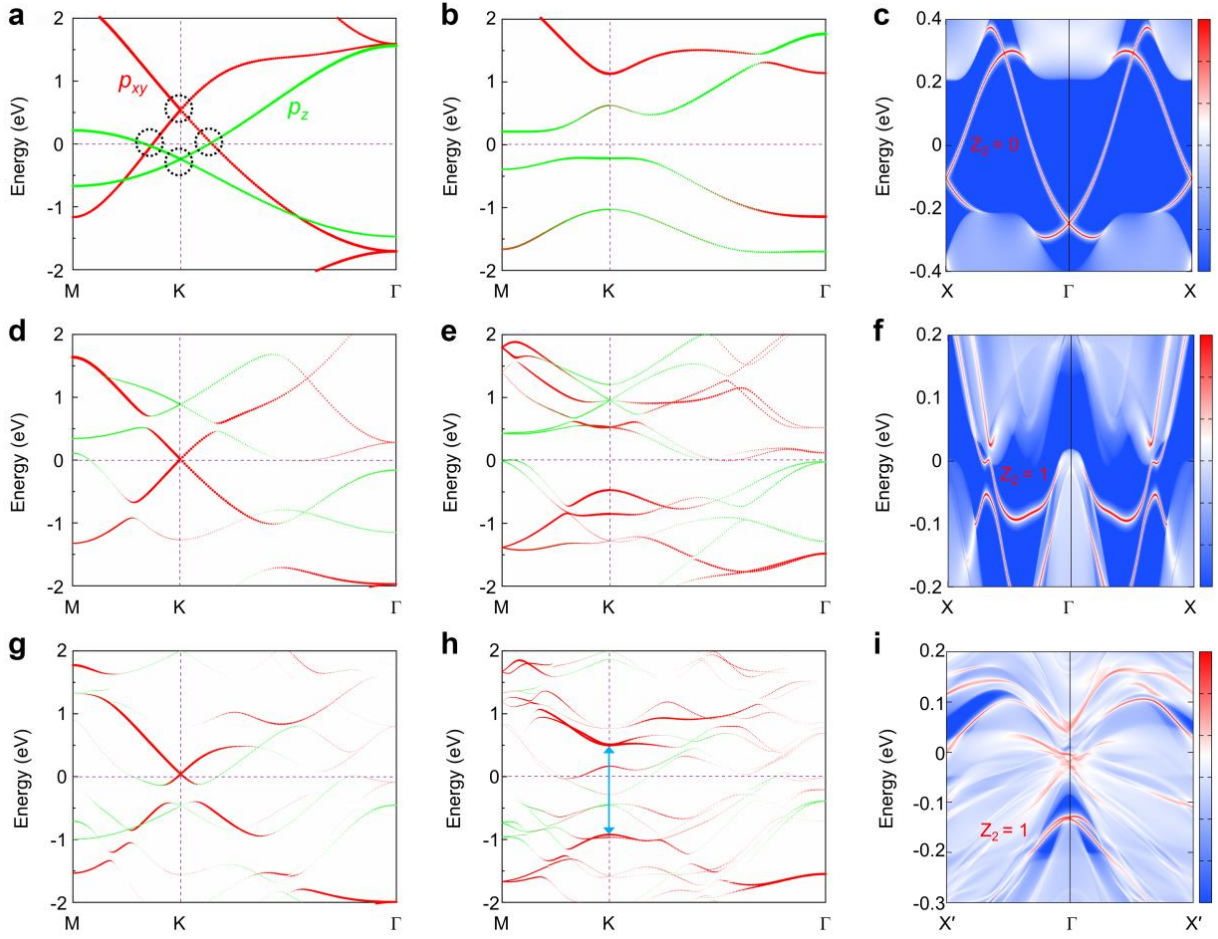

**Figure 2.** Calculated projected band structures of the ultraflat bismuthene. (a-c) Projected band structures for freestanding bismuthene without (a) and with (b) SOC, and edge states (c) for semi-infinite freestanding bismuthene with a zigzag-type termination. (d-f) Projected band structures for bismuthene/1L-Ag(111) without (d) and with (e) SOC, and edge states (f) for semi-infinite bismuthene/1L-Ag(111) with a zigzag-type termination. (g-i) Projected band structures for bismuthene/3L-Ag(111) without (g) and with (h) SOC, and edge states (i) for semi-infinite bismuthene/3L-Ag(111) with a zigzag-type termination. Red (green) dotted lines correspond to the $p_{xy}$ ($p_z$) orbitals of Bi atoms in the band structures. The color bars in panel (c), (f), (i) represent the density of states (DOS) from low (blue) to high (red). X′ = $0.23\pi/L_0$, where $L_0$ is the lattice periodicity.

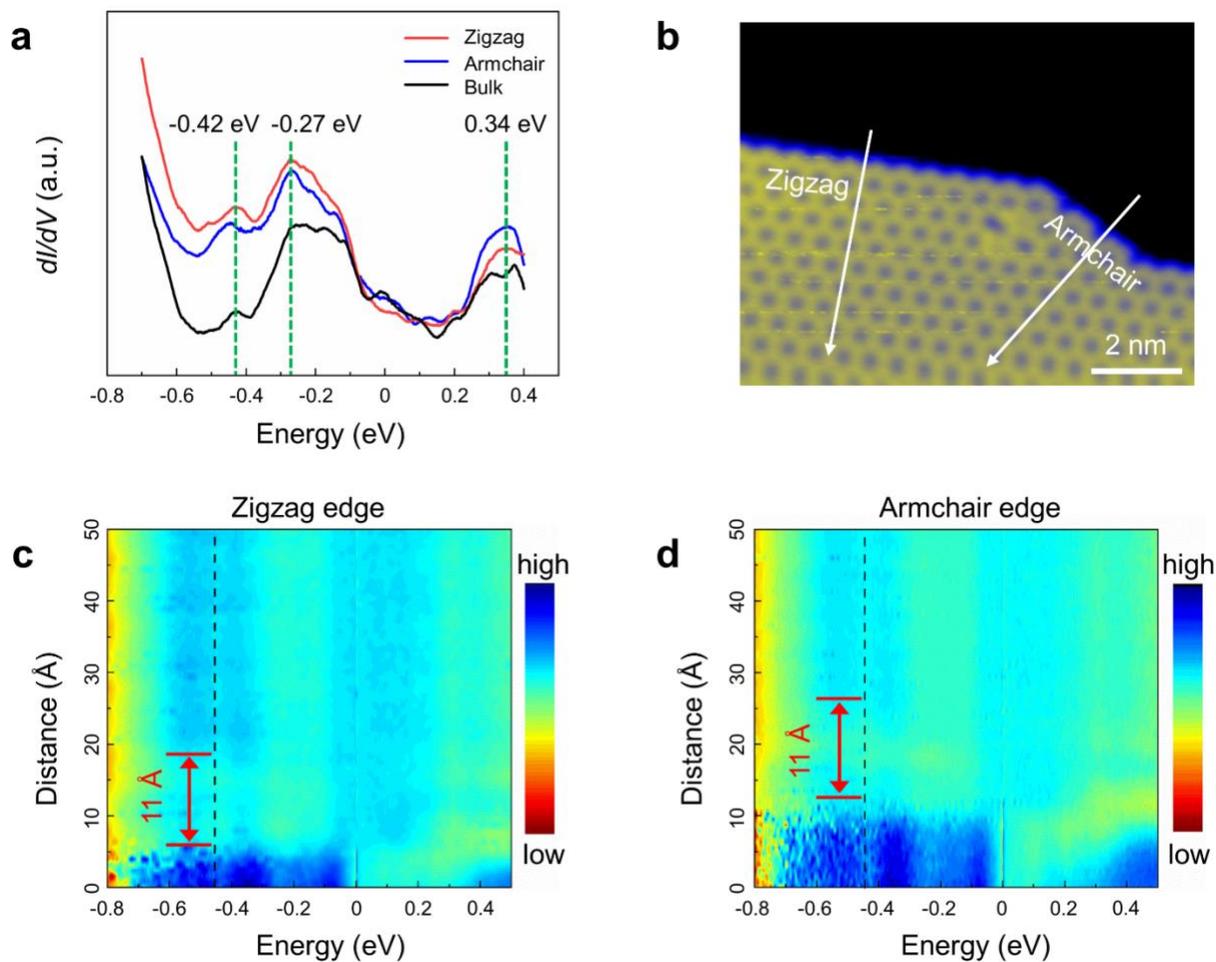

**Figure 3.** Topological edge states of the ultraflat bismuthene. (a) *dI/dV* curves taken at the zigzag edge, armchair edge and the central part of the ultraflat bismuthene island, respectively. Three obvious peaks are marked by green dashed lines. (b) STM image shows the co-existence of zigzag edge and armchair edge of a bismuthene island, where the white arrows indicating the trajectory of the line mappings ($V_s$ = 0.1 V, 8 × 10 nm$^2$). (c, d) *dI/dV* line mappings across the zigzag edge and armchair edge of the ultraflat bismuthene, respectively.

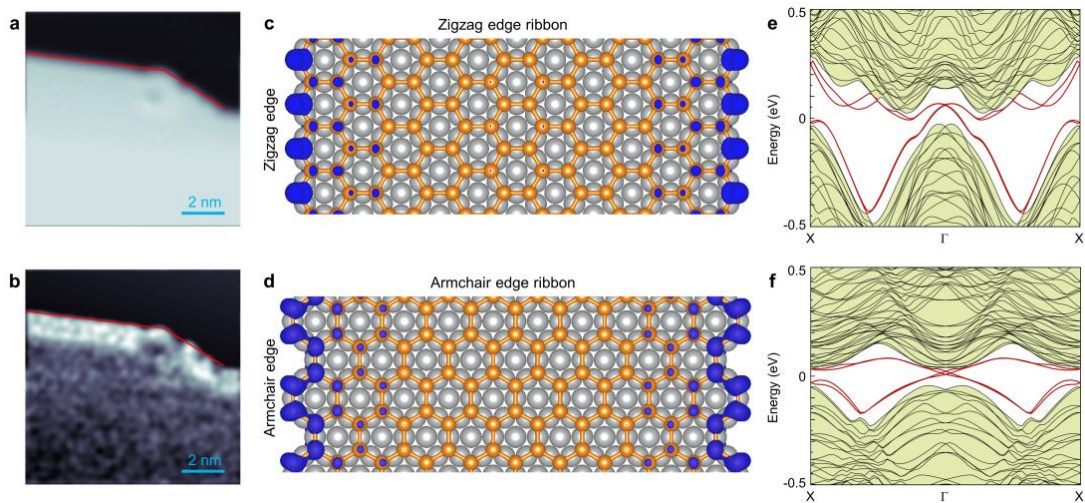

**Figure 4.** Spatial distribution of topological edge states in the bismuthene. (a, b) STM image (a) and the corresponding *dI/dV* mapping (b) of the bismuthene, where the red line is superimposed to highlight the spatial location of the edges ($V_s$ = -0.4 V, 10 × 10 nm$^2$). (c, d) Charge density distribution of the edge states in the zigzag-terminated nanoribbon with a width of 55.8 Å (c) and armchair-terminated nanoribbon with a width of 56.8 Å (d). (e, f) Calculated band structures of the zigzag- (e) and armchair-terminated nanoribbon (f), in which the bulk bands and helical edge states are represented by the shadow and red lines, respectively.

**Supplementary Figures and Notes.**

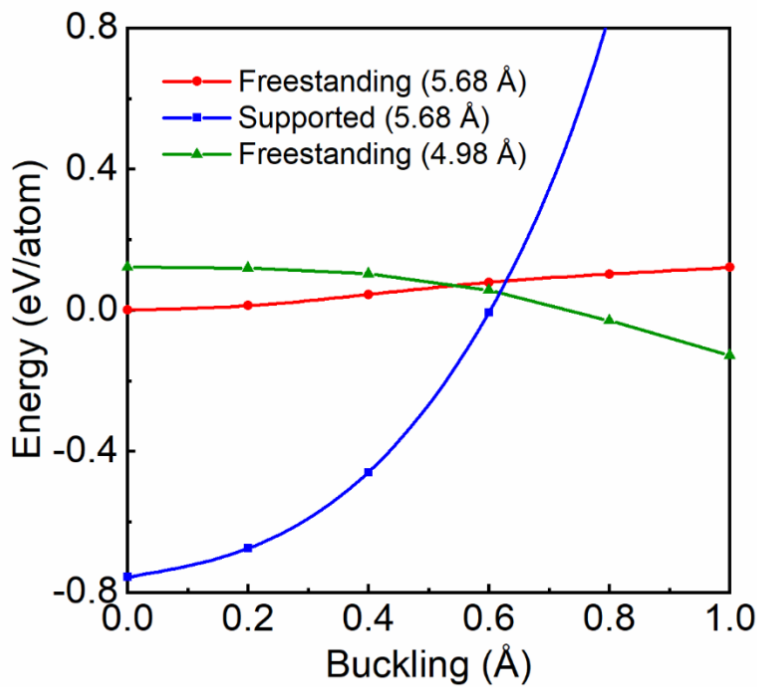

**Figure S1.** Total energy per Bi atom as a function of the buckling for the freestanding (with 5.68 Å and 4.98 Å lattice constants, respectively) and Ag(111)-supported bismuthene (with 5.68 Å lattice constant).

Previous studies suggest that buckled configuration plays an important role in stabilizing structure. The ultraflat bismuthene on Ag(111) features planar structure with no buckling. To further understanding the underlying mechanism, we calculated the total energy per Bi atom as a function of the buckling for the freestanding bismuthene monolayer with the lattice constant of 4.98 Å for comparison, as shown in Figure S1. Notably, the bismuthene monolayer with the lattice constant of 4.98 Å is expected to be a buckled configuration, which is indeed consistent with that a smaller buckling is preferred in a stretched lattice. However, in our experiment, the synthesized bismuthene features a larger lattice constant (5.7 Å). At such large lattice constant, a planar rather than buckled structure is preferred.

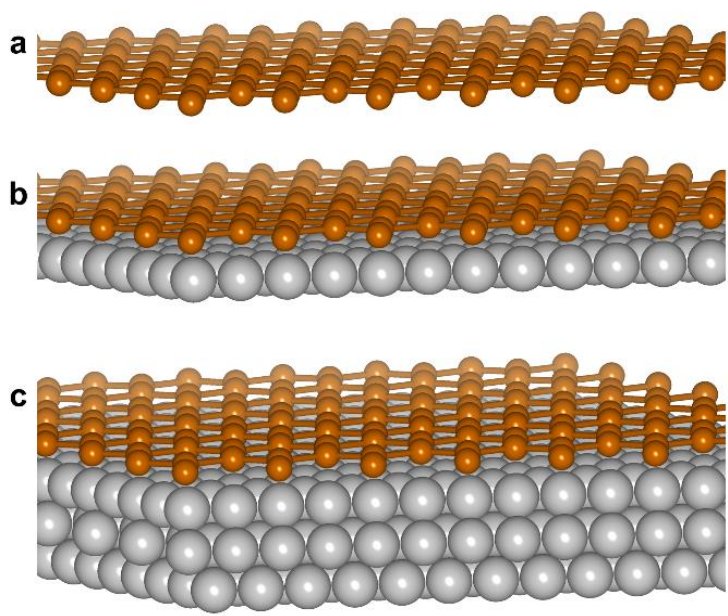

**Figure S2.** Atomic structures for (a) the freestanding bismuthene, (b) bismuthene/1L-Ag(111) and (c) bismuthene/3L-Ag(111), respectively. Bi (Ag) atoms are represented by orange (gray) balls.

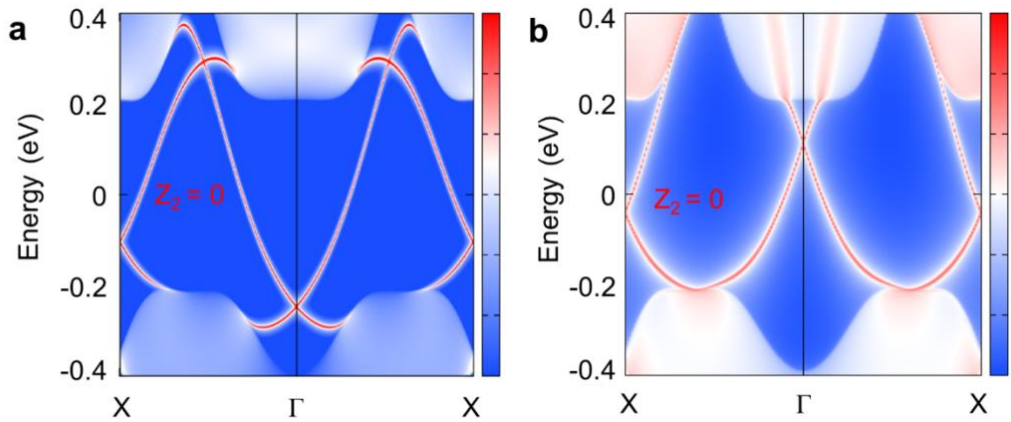

**Figure S3.** Projected band structures for semi-infinite freestanding bismuthene with a zigzag-type termination along the (100) direction (a) and (010) direction (b), respectively. The color bars represent the density of states (DOS) from low (blue) to high (red).

In Figure 2c (here Figure S3a), the edge states along the (100) direction degenerate in the path Γ-X. However, these degenerate points are occasional and not protected by symmetry. They cannot be opened to obtain an edge state within the energy gap. As we can see that there is no degenerate point of the edge states along the (010) direction in the path Γ-X (Figure S3b), making it a topological trivial phase.

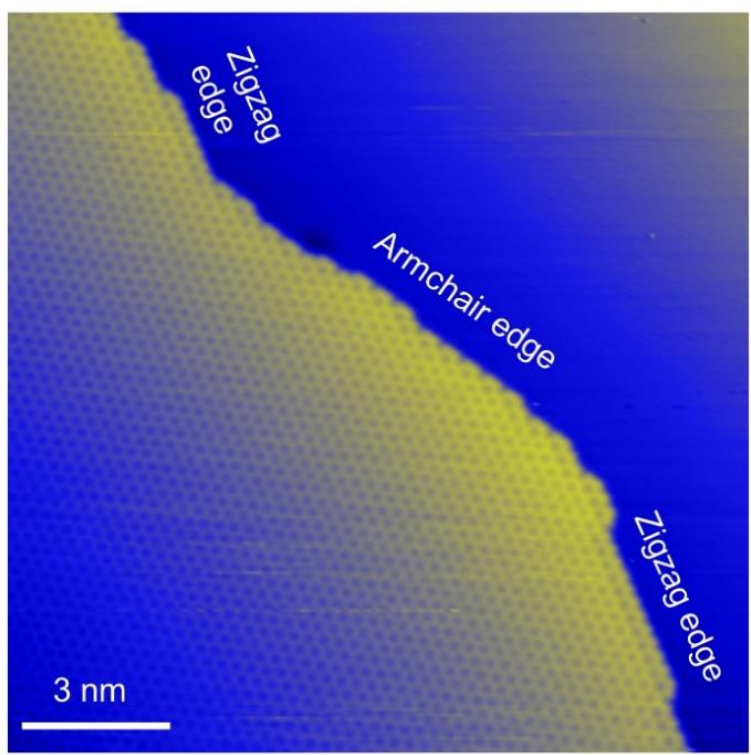

**Figure S4.** High resolution STM image with typical zigzag edges and armchair edges. ($V_s$ = -0.4 V, 15 × 15 nm$^2$). This is different from graphene and the other 2D materials, where the zigzag edges are mostly observed.

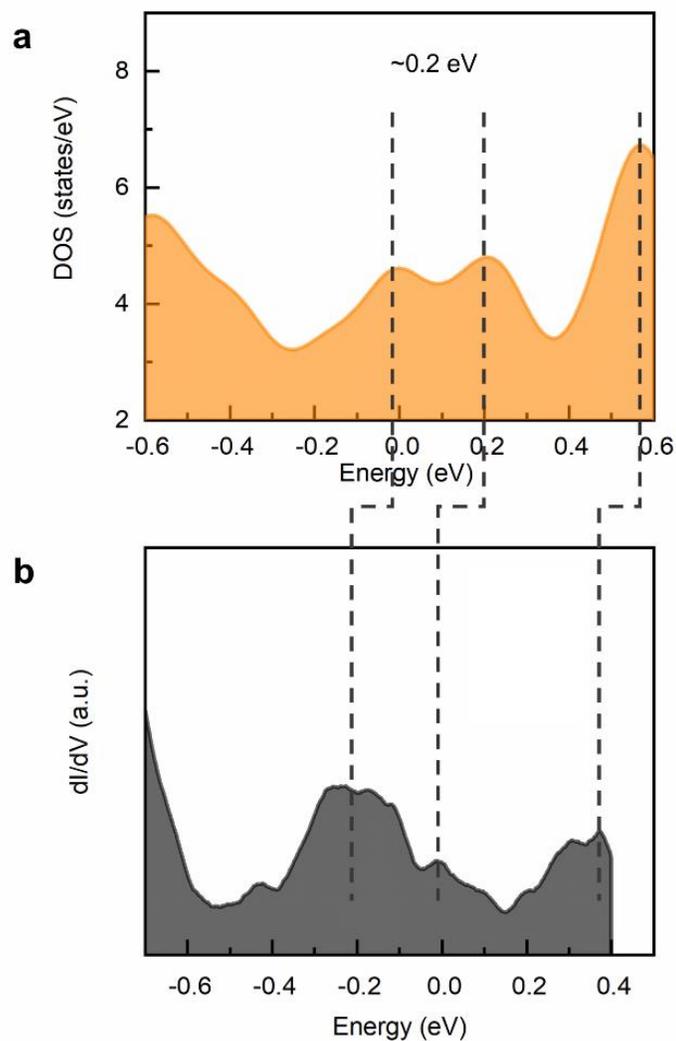

**Figure S5.** Comparison of calculated bulk DOS (a) for the bismuthene/3L-Ag(111) structure and experimental STS data (b).

The calculated bulk DOS has a similar shape comparing with the experimental STS, but with a slightly higher value than the experimental STS of about 0.2 eV. Such minor discrepancy is caused by the GGA in DFT calculations, which has also been reported in other systems, *e.g.*, ultraflat stanene on Cu(111).[1]

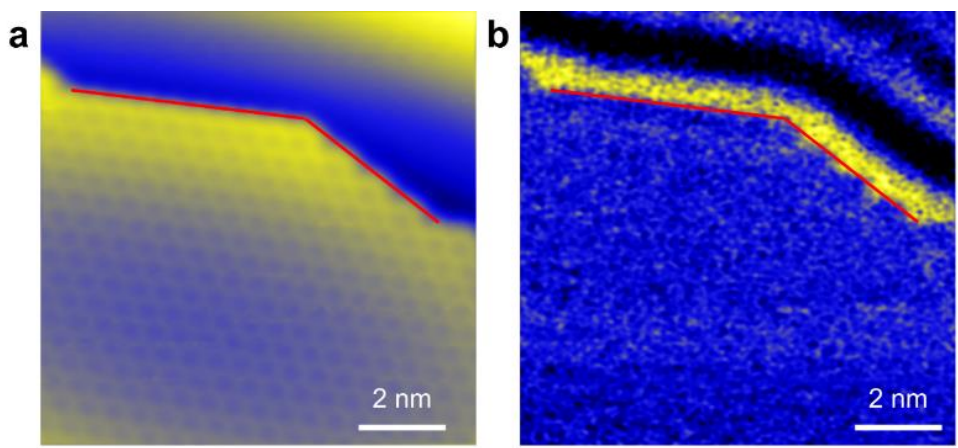

**Figure S6.** STM image (a) and *dI/dV* mapping (b) of the bismuthene with typical zigzag and armchair edges ($V_s$ = 0.3 V, 10 × 10 nm$^2$), where the red line is superimposed to highlight these edges.

It is obvious that the higher DOS in Figure S5b is indeed contributed by the underlying Ag surface rather than the edges of bismuthene.

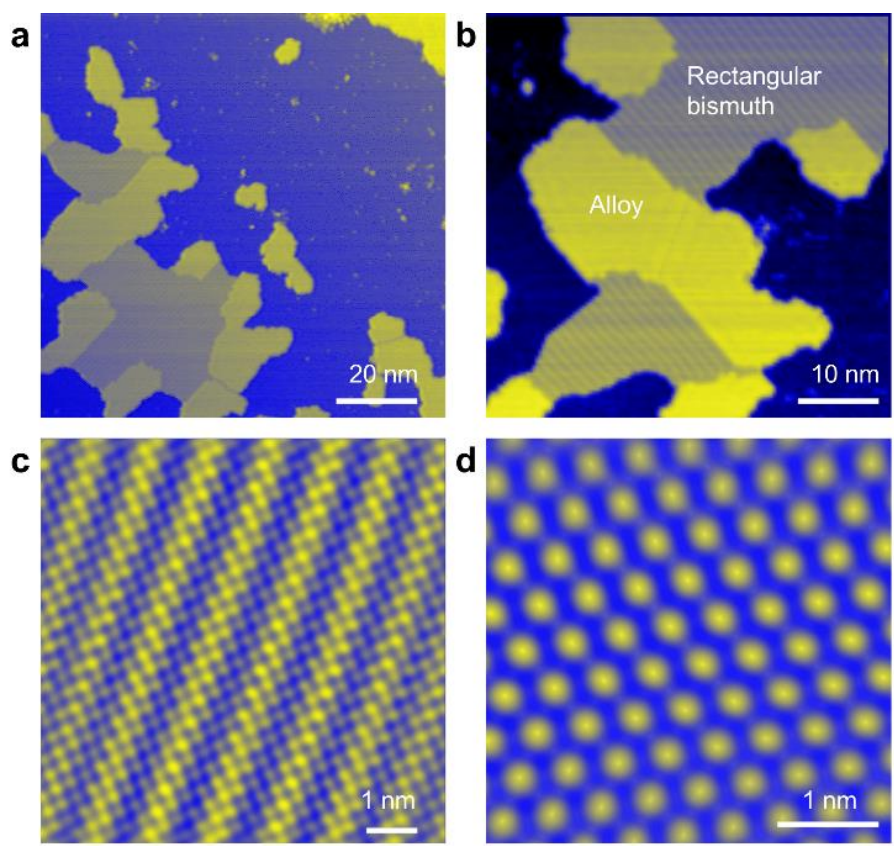

**Figure S7.** Growth behavior of bismuth on Ag(111) at higher substrate temperatures. Large-scale (a, b) and high-resolution (c, d) STM images of an allotrope of bismuth with rectangular symmetry and Bi-Ag alloy. (a-d: $V_s$ = 100 mV, 100 × 100 nm$^2$; $V_s$ = 100 mV, 50 × 50 nm$^2$; $V_s$ = 100 mV, 8 × 8 nm$^2$; $V_s$ = 10 mV, 4 × 4 nm$^2$)